\documentclass[twoside]{article}
\usepackage{qic,epsfig}
\usepackage{multirow}

\begin{document}
\setlength{\textheight}{8.0truein}    

\runninghead{Efficient rate-adaptive reconciliation for CV-QKD protocol}
            {Xiangyu Wang, Yichen Zhang, Zhengyu Li, Bingjie Xu, Song Yu, Hong Guo}

\normalsize\textlineskip
\thispagestyle{empty}
\setcounter{page}{1}

\vspace*{0.88truein}

\fpage{1}
\centerline{\bf
EFFICIENT RATE-ADAPTIVE RECONCILIATION}
\vspace*{0.035truein}
\centerline{\bf FOR CONTINUOUS-VARIABLE QUANTUM KEY DISTRIBUTION}
\vspace*{0.37truein}
\centerline{\footnotesize
XIANGYU WANG, YICHEN ZHANG, SONG YU\footnote{yusong@bupt.edu.cn}
}
\vspace*{0.015truein}
\centerline{\footnotesize\it State Key Laboratory of Information Photonics and Optical Communications, Beijing University
}
\baselineskip=10pt
\centerline{\footnotesize\it of Posts and Telecommunications, No. 10, Xitucheng Road, Beijing 100876, China
}
\vspace*{10pt}
\centerline{\footnotesize BINGJIE XU}
\vspace*{0.015truein}
\centerline{\footnotesize\it Science and Technology on Communication Security Laboratory, Institute of Southwestern}
\baselineskip=10pt
\centerline{\footnotesize\it Communication, No. 6, Chuangye Road, Chengdu 610041, China}
\vspace*{10pt}
\centerline{\footnotesize ZHENGYU LI, HONG GUO}
\vspace*{0.015truein}
\centerline{\footnotesize\it State Key Laboratory of Advanced Optical Communication Systems and Networks, School of Electronics}
\baselineskip=10pt
\centerline{\footnotesize\it Engineering and Computer Science, Center for Quantum Information Technology, Center for Computational}
\baselineskip=10pt
\centerline{\footnotesize\it Science and Engineering, Peking University, No. 5, Yiheyuan Road, Beijing 100871, China}
\vspace*{0.225truein}
\publisher{(received date)}{(revised date)}

\vspace*{0.21truein}
\abstracts{
Information reconciliation protocol has a significant effect on the secret key rate and maximal transmission distance of continuous-variable quantum key distribution (CV-QKD) systems. We propose an efficient rate-adaptive reconciliation protocol suitable for practical CV-QKD systems with time-varying quantum channel. This protocol changes the code rate of multi-edge type low density parity check codes, by puncturing (increasing the code rate) and shortening (decreasing the code rate) techniques, to enlarge the correctable signal-to-noise ratios regime, thus improves the overall reconciliation efficiency comparing to the original fixed rate reconciliation protocol. We verify our rate-adaptive reconciliation protocol with three typical code rate, i.e., 0.1, 0.05 and 0.02, the reconciliation efficiency keep around 93.5\%, 95.4\% and 96.4\% for different signal-to-noise ratios, which shows the potential of implementing high-performance CV-QKD systems using single code rate matrix.
}{}{}

\vspace*{10pt}
\vspace*{3pt}
\communicate{to be filled by the Editorial}

\vspace*{1pt}\textlineskip	
\section{Introduction}	        
\vspace*{-0.5pt}
\noindent
Quantum key distribution (QKD) \cite{QKD1, QKD2} is one of the most practical quantum information technologies, which allows the legitimate communication parties, Alice and Bob, to share secret keys. QKD mainly contains discrete-variable (DV) protocols \cite{DVQKD} and continuous-variable (CV) protocols \cite{CVQKD1, CVQKD2,CVMDIQKD1,CVMDIQKD2}. A DV-QKD protocol encodes key information on discrete Hilbert space, {\it i.e.} the polarization of a single photon. A CV-QKD protocol encodes key information on continuous variables, such as the quadratures of coherent states \cite{Gau1, Gau2, Gau3, Gau4}, which can directly use the standard telecommunication technologies. Thus CV-QKD protocol has more potential advantages in practical applications. In the Gaussian-modulated coherent state (GMCS) CV-QKD protocol \cite{Gau1}, Alice prepares GMCS and sends them to Bob through quantum channel. Then Bob randomly measures one of the quadratures with homodyne detector or both quadratures with heterodyne detector. Finally Bob informs Alice the quadrature he measures. After above processes, Alice and Bob share related Gaussian variables which are used to extract secret keys by post-processing algorithms.

The post-processing of CV-QKD protocol contains four main steps: base sifting, parameter estimation \cite{PE1, PE2}, information reconciliation \cite{IR3, IR2, IR1, IR4} and privacy amplification \cite{PA1, PA2}. The main bottleneck is information reconciliation. In GMCS CV-QKD protocol, the raw keys of Alice and Bob are Gaussian variables. Currently there are mainly two kinds of applied reconciliation methods to extract binary information from Gaussian variables, which are slice reconciliation \cite{slice1, slice2} and multidimensional reconciliation \cite{MD1, MD2}. Slice reconciliation is more suitable for the short-distance system, while multidimensional reconciliation can be used in the long-distance system. Multi-edge type low density parity check codes (MET-LDPC) \cite{METLDPC,CODE} are the generalization of LDPC codes \cite{CODE, LDPC}, which are used in multidimensional reconciliation, because its performance is close to the Shannon's limit especially at low signal-to-noise ratio (SNR) regime.

Currently, for low SNR, one can reach high reconciliation efficiency when using multidimensional reconciliation and MET-LDPC codes. However the codes are just applicable to some specific SNR. When the practical SNR differs from the code's optimal suitable SNR, the reconciliation efficiency will be decreased. Practically, due to the imperfect of optical sources or other factors, quantum channel is a time-varying channel whose SNR is changing over time. Finite number of MET-LDPC codes can not support the fully practical application. And it is not realistic and over-complex to use many different codes for each different practical SNRs.

To solve this problem, we adopt the rate-adaptive principle \cite{DVSP} and propose an efficient rate-adaptive reconciliation protocol which changes the code rate of MET-LDPC code to adapt the SNR of time-varying channel in CV-QKD system. The security of rate-adaptive principle is proven in \cite{SP, SecSP}. The performance of rate-adaptive can be further improved by some methods as mentioned in \cite{Blind,HIGHspeed,SymBlind}. This protocol is implemented by adding punctured bits whose value between Alice and Bob are unrelated and shortened bits whose value between Alice and Bob are the same into the raw keys. By adding punctured bits, the SNR between Alice and Bob's data increases, which will relatively increase the code rate. While by adding shortened bits, the SNR decreases, which will relatively decrease the code rate. The positions of punctured bits and shortened bits are randomly selected by Bob, which will be informed to Alice. Thus the rate-adaptive reconciliation protocol could keep high reconciliation efficiency within a range of SNR, which will increase the key rate and transmission distance of CV-QKD systems.

The paper is organized as follows: in section 2, we introduce information reconciliation of Gaussian variables based on multidimensional reconciliation and MET-LDPC codes. In section 3, we describe a rate-adaptive reconciliation protocol for CV-QKD system. The performance of the protocol is shown and analyzed in section 4. Finally, we conclude this paper in section 5.

\section{Information Reconciliation of Gaussian Variables Protocol}
\begin{figure} [htbp]
\vspace*{13pt}
\centerline{\psfig{file=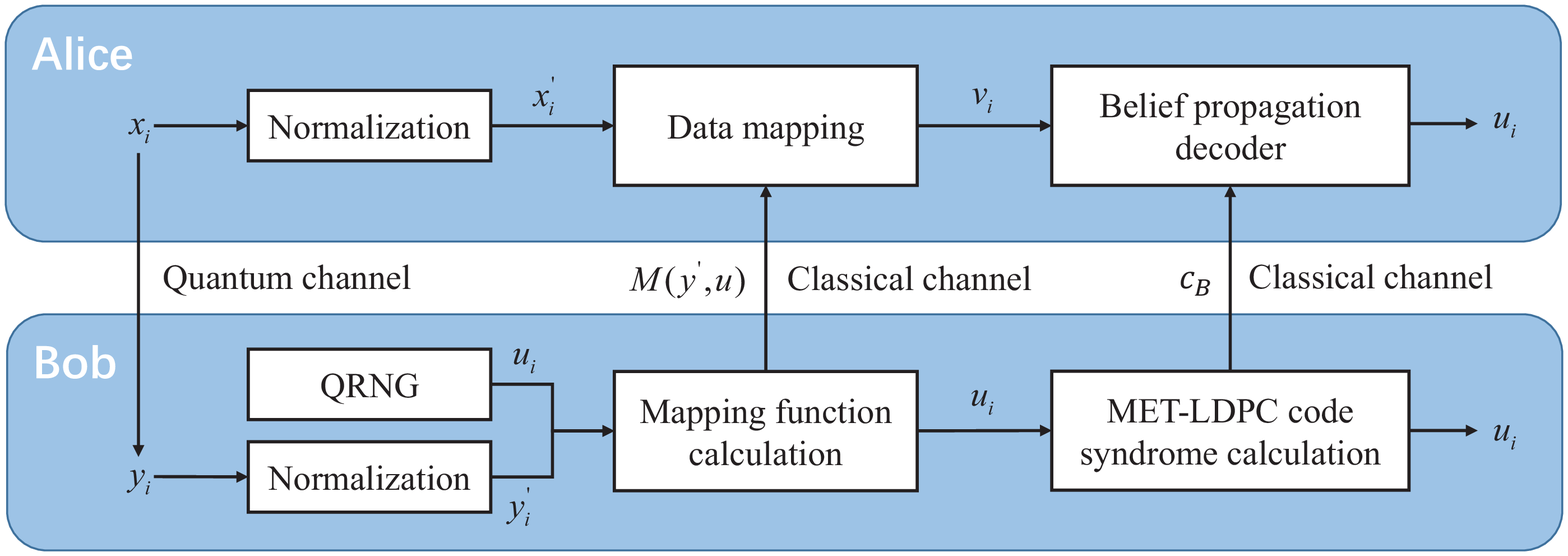, width=14.2cm}} 
\vspace*{13pt}
\fcaption{\label{informationreconciliation}
The principle of information reconciliation of CV-QKD protocol. First of all, Alice and Bob share related Gaussian variables ($x_{i}$ and $y_{i}$) through quantum channel. Then they use classical channel send some side information ($M(y', u)$ and $c_{B}$). Finally, they can share a common string ($u_{i}$) at a high probability based on these side information. QRNG: quantum random number generator.}
\end{figure}

\noindent
Information reconciliation has a significant effect on the secret key rate and transmission distance of CV-QKD systems. The information reconciliation of CV-QKD is divided into two parts. First, Alice and Bob use multidimensional reconciliation to construct a virtual binary input additive white Gaussian noise channel through rotating Gaussian variables. Second, Alice and Bob correct all the errors between their sequences based on MET-LDPC codes. For long-distance CV-QKD protocols, the reverse reconciliation \cite{RR} protocol is required, which takes Bob's sequence as target keys and corrects Alice's. Generally, before the information reconciliation step, Alice and Bob need to estimate the SNR of the quantum channel. The principle of information reconciliation of CV-QKD protocol is shown in Fig.~\ref{informationreconciliation}. Reconciliation efficiency is a significant parameter to evaluate the performance of the information reconciliation step. In CV-QKD protocols, when using multidimensional reconciliation, it is defined as follows:
\noindent
\begin{equation}
\beta=\frac{R}{C}
\,, \label{Reconciliationefficiency}
\end{equation}
where $R$ is the rate of MET-LDPC code, $C$ is the classical capacity of the quantum channel, which is $C=\frac{1}{2}log_{2}(1+SNR)$ for Gaussian variables \cite{MD1}.

Let $x$ and $y$ be the Gaussian variables of Alice's and Bob's. In GMCS CV-QKD protocol, the quantum channel is a additive white Gaussian noise channel. This means that one has $y=tx+z$, where $t$ is related with the channel loss, $z$ is the channel noise, $x$ and $z$ follow Gaussian distribution. For the information reconciliation step, SNR is the most crucial factor, thus for simplicity, one can fix $t=1$. Then one has $y=x+z$. In reverse reconciliation protocol, one has $x=y+z'$. Obviously $z'$ follows Gaussian distribution.

We explain the process of multidimensional reconciliation \cite{MD1} in details. The Gaussian variables $x$ and $y$ are two $n$-dimensional real vectors. Alice and Bob randomly or sequentially recombine $d$ elements together. And $d$ is the dimension of multidimensional reconciliation, which is chosen as $d=8$ in our scheme. Then Alice and Bob normalize their Gaussian variables $x$ and $y$ to $x'$ and $y'$ respectively. Bob randomly chooses a vector $u\in\{-1/\sqrt{d}, 1/\sqrt{d}\}^{d}$, such that $u$ follows uniform distribution on the $d$-dimensional unit sphere. For the security of CV-QKD system, this process needs a quantum random number generator (QRNG) \cite{QRNG}. Then Bob calculates a mapping function $M(y', u)$ such that $M(y',u)\cdot y'=u$ and sends the function to Alice by a public classical authenticated channel. Alice uses this function to map her Gaussian variables to $v$ such that $v=M(y',u)\cdot x'$. After the above processes, Alice and Bob have constructed a virtual binary input additive white Gaussian noise channel with input $u$ (Bob) and output $v$ (Alice) in reverse reconciliation protocol.

MET-LDPC codes~\cite{METLDPC, CODE} are error correction codes, which have the performance close to Shannon's limit at low SNR. Generally, MET-LDPC codes are represented by their parity check matrices. The rows of the matrices represent check nodes and the columns represent variable nodes. The number of nonzero in each row (column) represents the degrees of the check (variable) node. One can use density evolution method to get the degree distribution of matrices and estimate the threshold of the codes \cite{DE}. The threshold of a code is defined as the maximum standard deviation of noise channel that can be corrected when the code length is infinite and the maximum number of iterations is large enough (the standard deviation of signal is fixed to 1). In other words, the minimum SNR of the channel that can be corrected is $1/threshold^2$ for the code. The degree distribution of MET-LDPC code is specified by a multivariate polynomial pair ($\nu(\textbf{r,x}),\mu(\textbf{x})$), where $\nu(\textbf{r,x})$ is associated to variable nodes and $\mu(\textbf{x})$ is associated to check nodes~\cite{METLDPC, CODE}. The multivariate polynomial pair is defined as follows~\cite{METLDPC}:
\begin{equation}
\nu(\textbf{r,x})=\sum\nu_{\mathbf{b,d}}\textbf{r}^{\mathbf{b}}\textbf{x}^{\mathbf{d}}\,,
\end{equation}
\begin{equation}
\mu(\textbf{x})=\sum\mu_{\mathbf{d}}\textbf{x}^{\mathbf{d}}\,,
\end{equation}
where $\mathbf{b}$, $\mathbf{d}$, $\mathbf{r}$, $\mathbf{x}$, $\nu_{\mathbf{b,d}}$ and $\mu_{\mathbf{d}}$ are defined as follows. Vector $\mathbf{b}$ denote different types of channels. Typically, $\mathbf{b}$ only has two values (0 or 1), one denotes the channel which transmits bit, the other denotes the channel which punctures bit. Vector $\mathbf{d}$ denote the multi-edge degree. Vector $\mathbf{r}$ denote variables corresponding to the different types of channels. Vector $\mathbf{x}$ denote variables. $\nu_{\mathbf{b,d}}$ and $\mu_{\mathbf{d}}$ are the probabilities of variable nodes of type ($\mathbf{b,d}$) and check nodes of type $\mathbf{d}$. The code rate of MET-LDPC code is given by
\begin{equation}
R=\sum\nu_{\mathbf{b,d}}-\sum\mu_{\mathbf{d}}\,,
\label{ldpccoderate}
\end{equation}

Since we only use one type of channel (the channel which transmits bits), the sum of $\nu_{\mathbf{b,d}}$ is 1 and the sum of $\mu_{\mathbf{d}}$ is $1-R$, where $R$ is the code rate of MET-LDPC code. This is different from normal LDPC codes because that the definitions are different. For normal LDPC codes, they are defined as the ratios of the number of edges that connect to the variable nodes and check nodes to the total number of edges. Since the ratios are respectively corresponding to variable nodes and check nodes, the sum of variable node probabilities and check node probabilities are 1. However, for MET-LDPC codes, suppose the codeword length is $N$, $\nu_{\mathbf{b,d}}N$ is the number of variable nodes with type ($\mathbf{b}$, $\mathbf{d}$) and $\mu_{\mathbf{d}}N$ is the number of check nodes with type $\mathbf{d}$. Since the total number of variable nodes is equal to $N$, the sum of $\nu_{\mathbf{b,d}}$ is $1$. However, the total number of check nodes is $(1-R)N$, thus the sum of $\mu_{\mathbf{d}}$ is $1-R$. We use $\textbf{r}^{\mathbf{b}}$ and $\textbf{x}^{\mathbf{d}}$ to denote $\prod r_{i}^{b_{i}}$ and $\prod x_{i}^{d_{i}}$, respectively. The coefficients $\nu_{\mathbf{b,d}}$ and $\mu_{\mathbf{d}}$, the multi-edge degree $\mathbf{d}$ of variable nodes and check nodes are constrained to ensure the number of edges for each type is the same whether connect to variable nodes or check nodes.

We obtain the degree distribution of rate 0.1 and 0.05 codes by density evolution. In the appendix A of Ref.~\cite{IR3}, the degree distribution of a rate 0.02 code is described. The detailed degree distribution we used in this work is shown in Table~\ref{DegreeD}.

\begin{table}[htbp]
\tcaption{The multivariate polynomial pair of MET-LDPC codes.}
\centerline{\footnotesize\smalllineskip
\newcommand{\tabincell}[2]{\begin{tabular}{@{}#1@{}}#2\end{tabular}}
  \centering
  \renewcommand\arraystretch{2}
\begin{tabular}{|c|l|c|}
  \hline
  Code rate & \multicolumn{1}{c|}{Degree distribution} & Threshold  \\
  \hline
  0.1 & \tabincell{l}{$\nu(\textbf{r,x})=0.0775r_{1}x_{1}^{2}x_{2}^{20}+0.0475r_{1}x_{1}^{3}x_{2}^{22}+0.875r_{1}x_{3}$\\
  $\mu(\textbf{x})=0.0025x_{1}^{11}+0.0225x_{1}^{12}+0.03x_{2}^{2}x_{3}+0.845x_{2}^{3}x_{3}$} & 2.541  \\
  \hline
  0.05 & \tabincell{l}{$\nu(\textbf{r,x})=0.04r_{1}x_{1}^{2}x_{2}^{34}+0.03r_{1}x_{1}^{3}x_{2}^{34}+0.93r_{1}x_{3}$ \\ $\mu(\textbf{x})=0.01x_{1}^{8}+0.01x_{1}^{9}+0.41x_{2}^{2}x_{3}+0.52x_{2}^{3}x_{3}$} & 3.674  \\
  \hline
  0.02 & \tabincell{l}{$\nu(\textbf{r,x})=0.0225r_{1}x_{1}^{2}x_{2}^{57}+0.0175r_{1}x_{1}^{3}x_{2}^{57}+0.96r_{1}x_{3}$ \\ $\mu(\textbf{x})=0.010625x_{1}^{3}+0.009375x_{1}^{7}+0.6x_{2}^{2}x_{3}+0.36x_{2}^{3}x_{3}$} & 5.91\\
  \hline
\end{tabular}}
\label{DegreeD}
\end{table}

Based on the degree distribution, then one selects some methods to construct the parity check matrices. Typically, there are two kinds of methods, which are structured construction \cite{BIBD} and random construction \cite{PEG}. Technically, random construction has better error correction performance, which is more suitable for low SNR CV-QKD system. Thus, we choose progressive edge-growth method \cite{PEG} to construct parity check matrices, which is a good method of random construction.

Let $H$ be the parity check matrix, $m$ and $n$ be the number of rows and columns of $H$. The code rate of the matrix is $R=(n-m)/n$. In reverse reconciliation protocol, Bob calculates the syndromes $c_{B}$ of $u$, which is defined as $c_{B}=Hu^{T}$, and then he sends $c_{B}$ to Alice. Alice uses the belief propagation (BP) decoding algorithm to recover $u$ based on $c_{B}$, $v$ and $H$. Finally Alice and Bob share a common string $u$ with a certain probability. This probability depends on the difficulty of the error correction. Typically, the higher the reconciliation efficiency is, the greater the difficulty of the error correction is.

\section{Rate-adaptive Reconciliation Protocol}
\noindent
In this section,we present a rate-adaptive protocol for information reconciliation of CV-QKD. It can adapt the code rate to the time-varying quantum channel. This protocol is implemented by adding punctured bits and shortened bits \cite{DVSP} into the Gaussian variables of Alice and Bob, which is equivalent to change the rate of MET-LDPC code. The punctured bits increase the code rate and the shortened bits decrease the code rate. The detailed steps of the protocol are as follows:

\begin{figure} [htbp]
\vspace*{13pt}
\centerline{\psfig{file=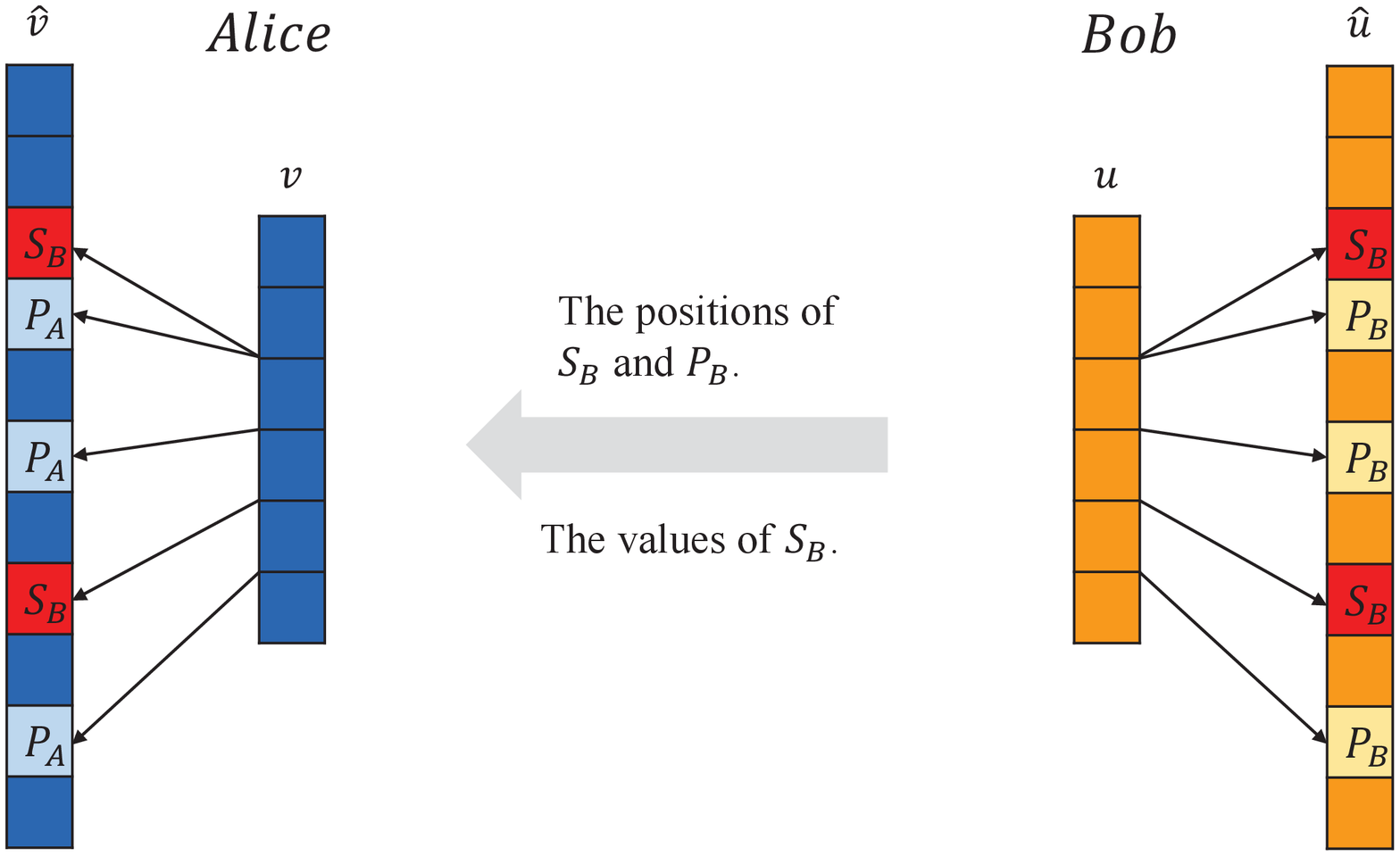, width=11cm}} 
\vspace*{13pt}
\fcaption{\label{constructnewstrings}
(color online) The process of Alice and Bob construct new strings for rate-adaptive reconciliation protocol. Firstly, Bob generates punctured bits $p_{B}$ and shortened bits $s_{B}$, and then randomly insert to string $u$. Secondly, Bob sends the positions of $p_{B}$ and $s_{B}$ and the values of $s_{B}$ to Alice. Thirdly, Alice reconstructs a new string $\hat{v}$ according to the message sent by Bob. The new string $\hat{v}$ contains string $v$, $p_{A}$ and $s_{A}$, where $s_{A}$=$s_{B}$.}
\end{figure}

{\it Step~1}: According to the practical SNR of the quantum channel, Alice and Bob calculate the optimal code rate. Then they select a good-performance original MET-LDPC code whose code rate is close to the optimal code rate. Generally, this original MET-LDPC code may not be the optimal code. The reasons are as followings:
\begin{alphlist}
\item Quantum channel is a time-varying channel whose SNR is fluctuant. Thus, the optimal code rate is not fixed.
\item We just have finite parity check matrices of MET-LDPC codes (the original code), which cannot support the full practical applications.
\end{alphlist}

Puncturing and shortening techniques are good ways to change the code rate. Punctured bits mean the data between Alice and Bob are completely unrelated, which increases the difficulty of error correction. Shortened bits are known to Alice and Bob, which helps the decoder to correct errors. Thus adding punctured bits increases the code rate and adding shortened bits decreases the code rate. Supposing the original code rate is $R^{o}=(n-m)/n$, $n$ is the length of the code and $m$ is the length of redundancy bits. As previously described, the code rate is Eq.~(\ref{ldpccoderate}). Actually, the two representations are equivalent. The quantity $\sum \nu_{\mathbf{b,d}}N$ is the number of variable nodes, {\it i.e.} the length of the code $n$, and the quantity $\sum \mu_{\mathbf{d}}N$ is the number of check nodes, {\it i.e.} the length of redundancy bits $m$. Because the original code only has one type of channel in our regime, such that $n=N$.
Thus, multiplying Eq.~\ref{ldpccoderate} by $N$ is $n-m$, then dividing it by $N$ is $R^{o}$. The puncturing technique changes the code rate by deleting some variable nodes and check nodes. Thus, the code rate will be changed to $R^{'}$ by adding punctured bits of length $p$.
\begin{equation}
R^{'}=\frac{(n-p)-(m-p)}{n-p}=\frac{n-m}{n-p}
\,,
\end{equation}
The shortening technique changes the code rate by deleting some variable nodes. Thus, the code rate will be changed to $R^{''}$ by adding shortened bits of length $s$.
\begin{equation}
R^{''}=\frac{(n-s)-m}{n-s}=\frac{n-m-s}{n-s}
\,,
\end{equation}
With the combination of puncturing and shortening techniques, the code rate will be changed to
\noindent
\begin{equation}
R=\frac{(n-p-s)-(m-p)}{n-p-s}=\frac{n-m-s}{n-p-s}
\,. \label{coderate}
\end{equation}

{\it Step~2}: Bob creates a new sequence $\hat{u}$ with length $n$:
\begin{equation}
u=\{u_{1},u_{2},\cdots\cdots,u_{n-p-s}\}\,,
\end{equation}
\begin{equation}
p_{B}=\{{p_{B}}_{1},{p_{B}}_{2},\cdots\cdots,{p_{B}}_{p}\}\,,
\end{equation}
\begin{equation}
s_{B}=\{{s_{B}}_{1},{s_{B}}_{2},\cdots\cdots,{s_{B}}_{s}\}\,,
\end{equation}
\begin{equation}
\widehat{u}=\{u_{1},u_{2},\cdots,{s_{B}}_{1},\cdots,{p_{B}}_{1},\cdots,{s_{B}}_{i},\cdots,{p_{B}}_{j},\cdots,u_{k},\cdots\}\,.
\end{equation}
where $u$ is the string for Bob's multidimensional reconciliation with length $n-p-s$, the sequences $p_{B}$ and $s_{B}$ represent Bob's punctured bits and shortened bits which are randomly generated by Bob with length $p$ and $s$, and $i\in{\{1,2,\cdots,s\}}$, $j\in{\{1,2,\cdots,p\}}$, $k\in{\{1,2,\cdots,n-p-s\}}$. The new string $\hat{u}$ is created by randomly inserting $p_{B}$ and $s_{B}$ into the string $u$. Then Bob calculates the syndrome of $\hat{u}$, such that $c(\hat{u})=H\hat{u}^{T}$. $H$ is the parity check matrix of MET-LDPC code. Finally Bob sends $c(\hat{u})$, $s$ shortened bits and the positions of punctured bits and shortened bits to Alice. The process is shown in the right of Fig.~\ref{constructnewstrings}.

{\it Step~3}: Alice receives the message sent by Bob. Then Alice constructs a new sequence $\hat{v}$ with length $n$:
\begin{equation}
v=\{v_{1},v_{2},\cdots\cdots,v_{n-p-s}\}\,,
\end{equation}
\begin{equation}
p_{A}=\{{p_{A}}_{1},{p_{A}}_{2},\cdots\cdots,{p_{A}}_{p}\}\,,
\end{equation}
\begin{equation}
\widehat{v}=\{v_{1},v_{2},\cdots,{s_{B}}_{1},\cdots,{p_{A}}_{1},\cdots,{s_{B}}_{i},\cdots,{p_{A}}_{j},\cdots,v_{k},\cdots\}\,.
\end{equation}
where $v$ is the sequence for Alice's multidimensional reconciliation with length $n-p-s$, the sequence $p_{A}$ represents Alice's punctured bits which is randomly generated by Alice, and $i\in{\{1,2,\cdots,s\}}$, $j\in{\{1,2,\cdots,p\}}$, $k\in{\{1,2,\cdots,n-p-s\}}$. The new string $\hat{v}$ and $\hat{u}$ have the same positions and length of punctured bits and shortened bits. And they also have the same value of shortened bits $s_{B}$. The process is shown in the left of Fig.~\ref{constructnewstrings}. Then Alice uses the belief propagation decoding algorithm or some other algorithm to recover $\hat{u}$. Finally Alice and Bob will share a common string.

{\it Example:} Supposing the practical SNR of quantum channel is 0.028, and we have a MET-LDPC code with rate 0.02. The code length $n$ is $10^{6}$ and the length of parity check bits $m$ is $980000$. The reconciliation efficiency can reach to 96.9\% when SNR is 0.029. According to Eq.~(\ref{Reconciliationefficiency}) we calculate that the optimal code rate is $0.0192$. According to Eq.~(\ref{coderate}), we can calculate that the length of shortened bits $s$ is $992$ and the length of punctured bits $p$ is $9008$. Thus, the reconciliation efficiency can be achieved around 96.38\%. We will show more results in the next section.

\section{Performance of the Protocol}
\noindent
In this section, we analysis the performance of the rate-adaptive reconciliation protocol. Reconciliation efficiency is one of the most important parameters to show the performance of the information reconciliation. Thus we mainly analyze the influence of the rate-adaptive reconciliation protocol on reconciliation efficiency. In the previous work on the postprocessing of CV-QKD protocol, for high SNRs (normally higher than 0.5), several work have been done to obtain and maintain high reconciliation efficiency \cite{HIGHBIT,NONBINLDPC}. In Ref.~\cite{HIGHBIT}, they use slice reconciliation achieve high reconciliation efficiencies at different SNRs by using different fixed-rate codes. In Ref.~\cite{NONBINLDPC}, they quantify the continuous variables and use non-binary LDPC codes to obtain excellent reconciliation efficiencies. And they also show that the reconciliation efficiencies can be maintained at a high level in a range of SNRs by varying the width of the reconciliation interval. Thus, we mainly focus on obtaining excellent reconciliation efficiencies at low SNRs and using rate-adaptive technique to optimize the efficiency in a range of SNRs.

For low SNRs (normally lower than 0.5), we obtain excellent reconciliation performance by combining multidimensional reconciliation and MET-LDPC codes. By using the aforementioned MET-LDPC codes, for code rates of 0.1, 0.05 and 0.02, we can get the reconciliation efficiencies to 93.95\%, 95.84\% and 96.99\% respectively. The results are better than those in~\cite{IR3}. The code length is $10^6$ in our system and their code length is $2^{20}$. The parity check matrix we used are constructed according to the degree distribution in strict, while theirs is not. And we substitute true random number for pseudo random number to select the positions of 1s in the parity check matrix, which makes the parity check matrix more random. Therefore, even though our code length is lower than theirs', we obtain better results. However, these results can be obtained only when the SNRs are 0.159, 0.075 and 0.029 respectively. Practically, quantum channel is a time-varying channel whose SNR may vary within a range. Thus, we propose a rate-adaptive reconciliation protocol in CV-QKD system, which can change the code rate to adapt the practical system. We show the detailed results of both reconciliation protocols in Table~\ref{RAresults}. The results of each data point in Table~\ref{RAresults} are the average of the 1000 simulation results.

\begin{table}[htbp]
\tcaption{The detailed results of the rate-adaptive reconciliation and reconciliation by original codes. $R^{o}$: the rate of original code. SNR: practical signal-to-noise ratio. s: the length of shortened bits. p: the length of punctured bits. R: the code rate after rate-adaptive. $\beta$: reconciliation efficiency of the rate-adaptive reconciliation. $\beta^{o}$: reconciliation efficiency of the original code.}
\centerline{\footnotesize\smalllineskip
\begin{tabular}{|c|c c c c c c || c c c c c c|}
\hline
{$R^{o}$}&{SNR} &{s} &{p} &{R} &{$\beta$} &{$\beta^{o}$} &{SNR} &{s} &{p} &{R} &{$\beta$} &{$\beta^{o}$}\\
\hline
\multirow{3}{*}{0.1}
& 0.143 &11080 &920 & 0.090 & 93.35\% &0\% & 0.163 &0 &19608 & 0.102 & 93.64\% &91.81\%\\
& 0.148 &7928 &2072 & 0.093 & 93.41\% &0\% & 0.169 &0 &47616 & 0.105 & 93.22\% &88.78\%\\
& 0.153 &4768 &3232 & 0.096 & 93.48\% &0\% & 0.176 &0 &82568 & 0.109 & 93.21\% &85.51\%\\
\hline
\multirow{3}{*}{0.05}
&0.069 &4656 &5344 & 0.0458 & 95.16\% &0\% & 0.077 &0 &23440 & 0.0512 & 95.68\% &93.44\%\\
&0.071 &3272 &6728 & 0.0472 & 95.39\% &0\% & 0.079 &0 &43976 & 0.0523 & 95.36\% &91.16\%\\
&0.073 &1984 &8016 & 0.0485 & 95.43\% &0\% & 0.081 &0 &65416 & 0.0535 & 95.22\% &88.99\%\\
\hline
\multirow{3}{*}{0.02}
&0.0277 &1200 &8800 & 0.0190 & 96.40\% &0\% & 0.0299 &0 &24392 & 0.0205 & 96.46\% &94.11\%\\
&0.0280 &992 &9008 & 0.0192 & 96.38\% &0\% & 0.0306 &0 &47616 & 0.0210 &  96.59\% &91.99\%\\
&0.0286 &592 &9408 & 0.0196 & 96.36\% &0\% & 0.0314 &0 &69768 & 0.0215 &  96.40\% &89.68\%\\
\hline
\end{tabular}}
\label{RAresults}
\end{table}

\begin{figure}[htpb]
\vspace*{13pt}
\begin{minipage}[t]{0.33\linewidth}
\centering
\includegraphics[width=\textwidth]{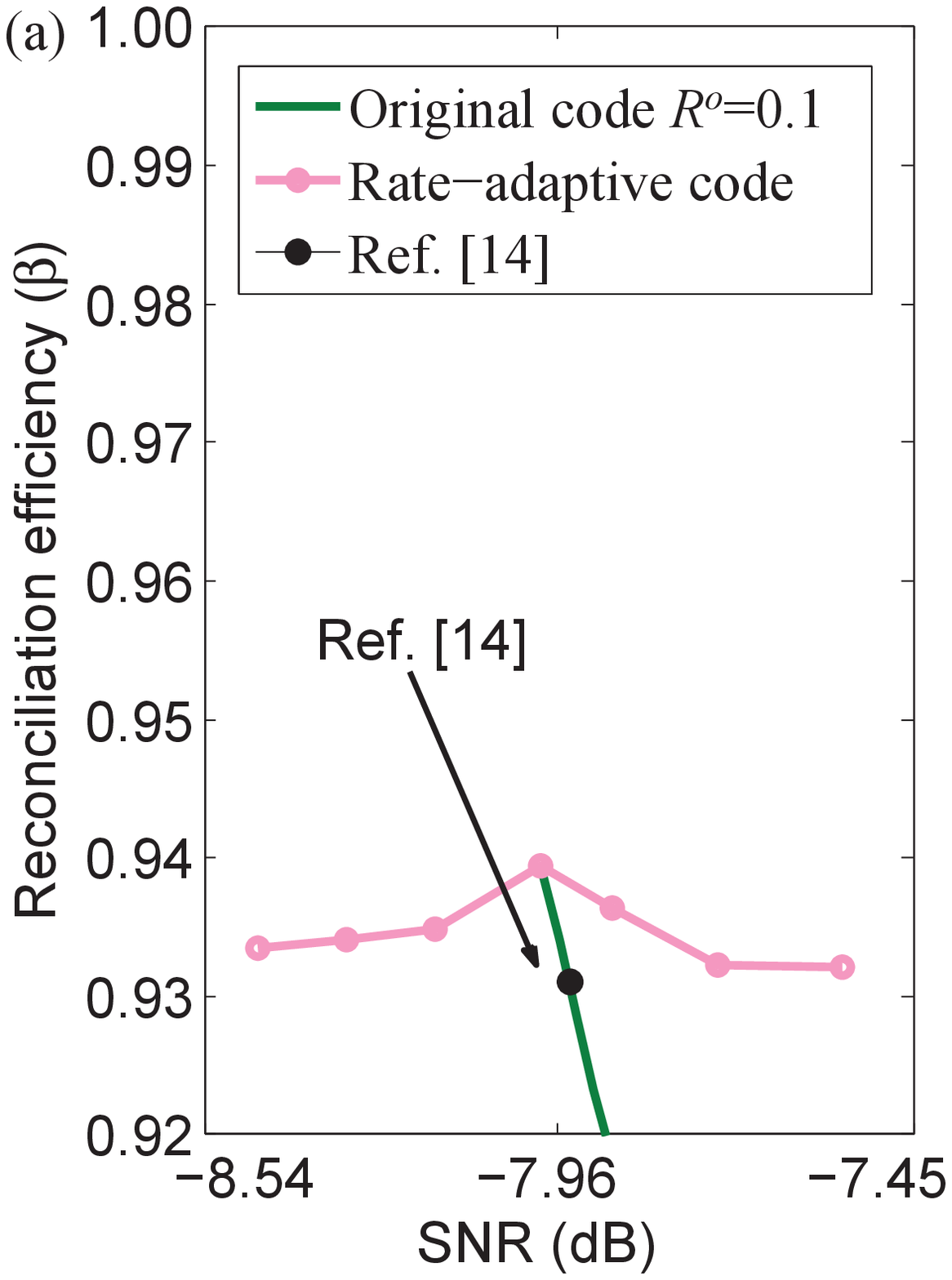}
\end{minipage}%
\begin{minipage}[t]{0.33\linewidth}
\centering
\includegraphics[width=\textwidth]{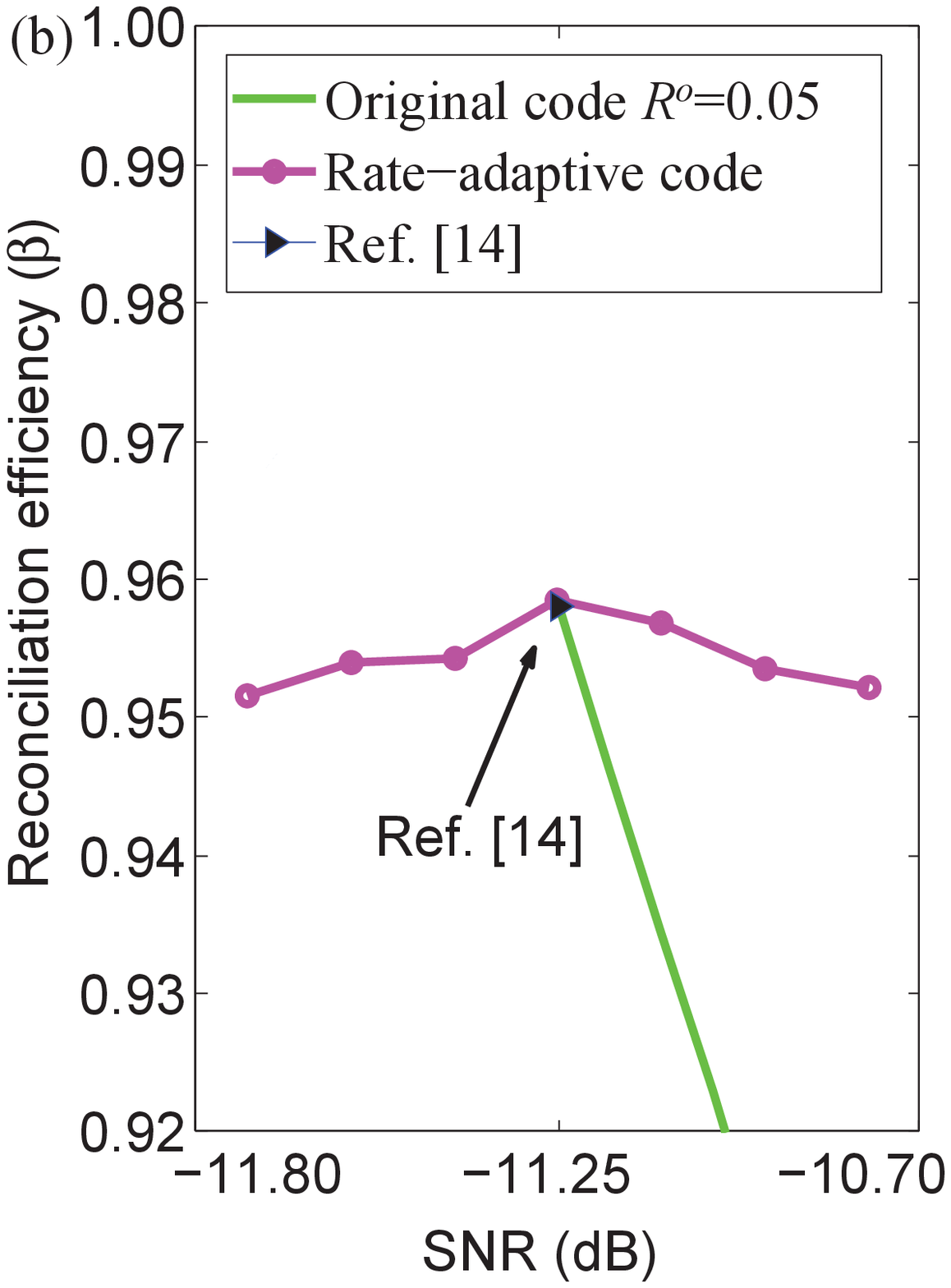}
\end{minipage}
\begin{minipage}[t]{0.33\linewidth}
\centering
\includegraphics[width=\textwidth]{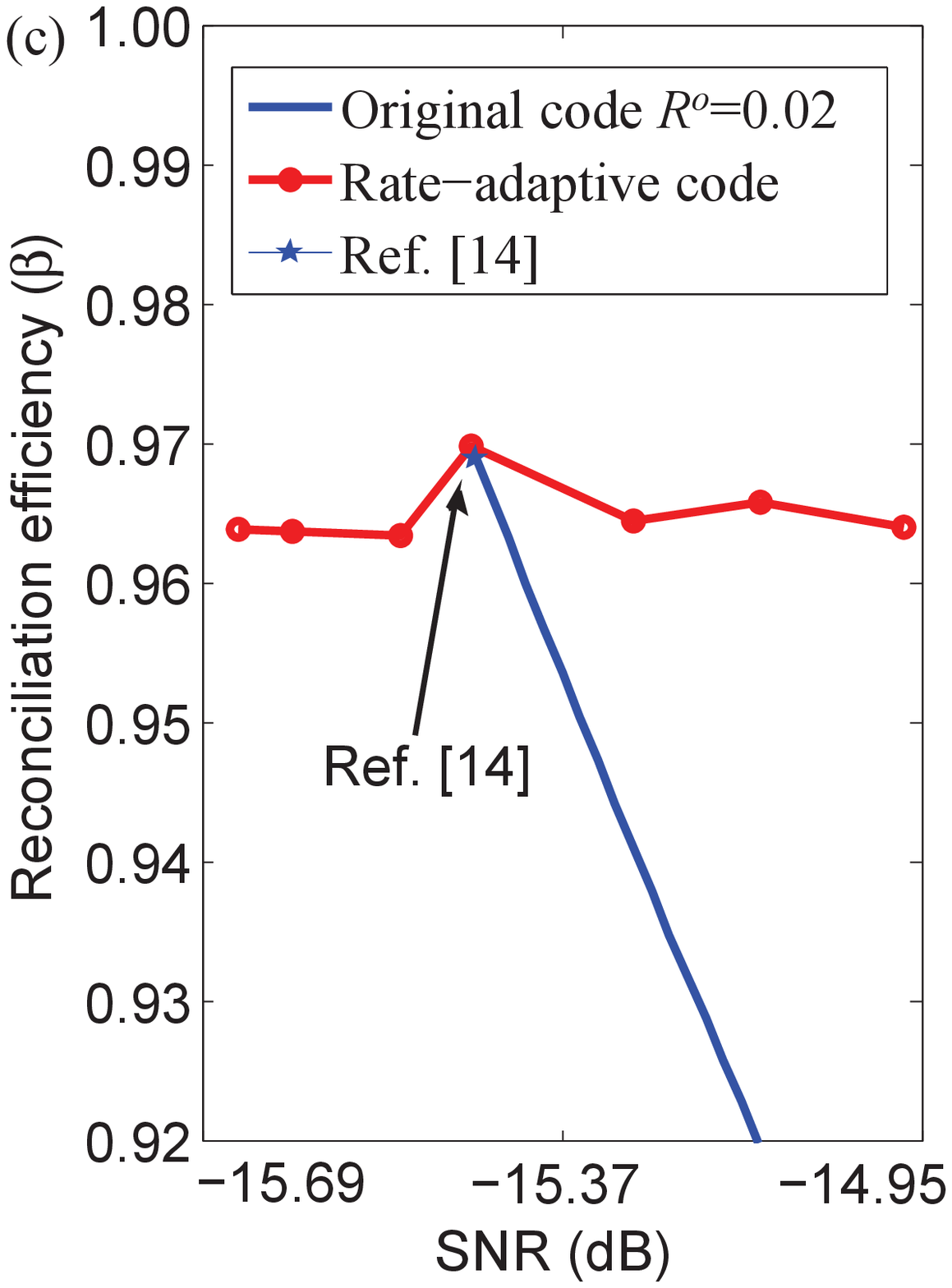}
\end{minipage}
\vspace*{13pt}
\fcaption{\label{002}
Performance comparison of reconciliation by original MET-LDPC codes and the rate-adaptive reconciliation protocol. The dark green line, light green line and blue line represent reconciliation efficiencies by using the original MET-LDPC code with code rate 0.1, 0.05 and 0.02 at different SNRs. The circle, triangle and star represent the reconciliation efficiency of code rate 0.1, 0.05 and 0.02 used in \cite{IR3}. The pink dots, purple dots and red dots represent reconciliation efficiencies of the rate-adaptive reconciliation protocol.}
\end{figure}

The original rates of MET-LDPC codes are 0.1, 0.05 and 0.02. According to the practical SNRs and the error correction performance of original codes, we use Eq.~(\ref{Reconciliationefficiency}) to calculate the optimal code rates. Then we use Eq.~(\ref{coderate}) to calculate the length of shortened bits $s$ and punctured bits $p$. Actually, $s$ and $p$ are not unique, as long as they satisfies Eq.~(\ref{coderate}). Although $s$ and $p$ are not unique, the ratio of $(s+p)/n$ has better not too big, otherwise the error correction performance will decrease. We show one set of results on each code rate in Table~\ref{RAresults}. As shown in Table~\ref{RAresults}, when the practical SNRs are less than the optimal SNR corresponding to the MET-LDPC code ( For the code rates of 0.1, 0.05 and 0.02, the optimal SNRs are 0.159, 0.075 and 0.029 respectively ), error correction will fail ( The 0\% reconciliation efficiency means that error correction fails ). When the practical SNRs are higher than the optimal SNR, reconciliation efficiency will be reduced. Both of the cases will decrease the key rates of CV-QKD system. However, no matter the practical SNRs are less than or higher than the optimal SNR, reconciliation efficiency of the rate-adaptive reconciliation is almost not decreased (keep around 93.5\%, 95.4\% and 96.4\% respectively). In Fig.~\ref{002}, We compare the performance of the rate-adaptive reconciliation protocol and reconciliation protocol by the original code used in \cite{IR3} under different SNRs. The reconciliation efficiencies of both protocols are obtained according to Eq.~(\ref{Reconciliationefficiency}).

As shown in Fig.~\ref{002}, for reconciliation by original codes used in \cite{IR3}, because the code rate is fixed, high reconciliation efficiency can be obtained only when the practical SNR corresponding to the optimal error correction performance of the MET-LDPC codes. Otherwise the reconciliation efficiency will be reduced quickly, even if the SNRs change in a very small range. However, because the code rate is changed according to the practical SNR, the rate-adaptive reconciliation can tolerate a certain fluctuation in SNRs and reconciliation efficiency is almost not reduced.

\section{Conclusion}
\noindent
We present an efficient rate-adaptive reconciliation protocol for continuous-variable quantum key distribution system. By using multidimensional reconciliation and multi-edge type low density parity check codes, we reach high reconciliation efficiency on specific signal-to-noise ratio. However, the signal-to-noise ratio of quantum channel may change over time even for a fixed distance, due to the varying environment. The protocol described in this paper can change the code rate according to the practical situation by puncturing and shortening techniques. Our results show that the rate-adaptive reconciliation keeps high efficiency in a range ($>$10\%) of signal-to-noise ratios, which will improve the robustness of practical continuous-variable quantum key distribution system.

\nonumsection{Acknowledgements}
\noindent
This work was supported in part by the National Basic Research Program of China (973 Pro-gram) under Grant 2014CB340102, in part by the National Natural Science Foundation under Grants 61531003, 61427813, 61401036, 61471051, 61501414, in part by Youth research and innovation program of BUPT (2015RC12).

\nonumsection{References}


\noindent

\begin{thebibliography}{000}
\bibitem{QKD1}
N. Gisin, G. Ribordy, W. Tittel and H. Zbinden (2002), {\it Quantum cryptography}, Rev. Mod. Phys. 74(1), 145.
\bibitem{QKD2}
V. Scarani, H. Bechmann-Pasquinucci, N. J. Cerf, M. Du\v{s}ek, N. L\"{u}tkenhaus and M. Peev (2009), {\it The security of practical quantum key distribution}, Rev. Mod. Phys. 81(3), 1301.
\bibitem{DVQKD}
C. H. Bennett and G. Brassard (1984), {\it Quantum cryptography: Public key distribution and coin tossing}, In Proceedings of International Conference on Computer System and Signal Processing, IEEE, 1984 pp. 175-179.
\bibitem{CVQKD1}
S. L. Braunstein and P. Van Loock (2005), {\it Quantum information with continuous variables}, Rev. Mod. Phys. 77(2), 513.
\bibitem{CVQKD2}
C. Weedbrook, S. Pirandola, R. Garc\'{\i}a-Patr\'{o}n, N. J. Cerf, T. C. Ralph, J. H. Shapiro and S. Lloyd (2012), {\it Gaussian quantum information}, Rev. Mod. Phys. 84(2), 621.
\bibitem{CVMDIQKD1}
S. Pirandola, C. Ottaviani, G. Spedalieri, C. Weedbrook, S. L. Braunstein, S. Lloyd, T. Gehring, S. J. Christian and U. L. Andersen (2015), {\it High-rate measurement-device-independent quantum cryptography}. Nat. Photon. 9(6), pp. 397-402.
\bibitem{CVMDIQKD2}
S. Pirandola, C. Ottaviani, G. Spedalieri, C. Weedbrook, S. L. Braunstein, S. Lloyd, T. Gehring, S. J. Christian and U. L. Andersen (2015), {\it Reply to'Discrete and continuous variables for measurement-device-independent quantum cryptography'}, Nat. Photon. 9(12), pp. 773-775.
\bibitem{Gau1}
F. Grosshans and P. Grangier (2002), {\it Continuous variable quantum cryptography using coherent states}, Phys. Rev. Lett. 88(5), 057902.
\bibitem{Gau2}
C. Weedbrook, A. M. Lance, W. P. Bowen, T. Symul, T. C. Ralph and P. K. Lam (2004), {\it Quantum cryptography without switching}, Phys. Rev. Lett. 93(17), 170504.
\bibitem{Gau3}
S. Pirandola, S. Mancini, S. Lloyd and S. L. Braunstein (2008), {\it Continuous-variable quantum cryptography using two-way quantum communication}, Nat. Phys. 4(9), pp. 726-730.
\bibitem{Gau4}
Z. Li, Y. C. Zhang, F. Xu, X. Peng and H. Guo (2014), {\it Continuous-variable measurement-device-independent quantum key distribution}, Phys. Rev. A, 89(5), 052301.
\bibitem{PE1}
A. Leverrier, F. Grosshans and P. Grangier (2010), {\it Finite-size analysis of a continuous-variable quantum key distribution}, Phys. Rev. A, 81(6), 062343.
\bibitem{PE2}
P. Jouguet, S. Kunz-Jacques, E. Diamanti and A. Leverrier (2012), {\it Analysis of imperfections in practical continuous-variable quantum key distribution}, Phys. Rev. A, 86(3), 032309.
\bibitem{IR3}
P. Jouguet, S. Kunz-Jacques and A. Leverrier (2011), {\it Long-distance continuous-variable quantum key distribution with a Gaussian modulation}, Phys. Rev. A, 84(6), 062317.
\bibitem{IR2}
G. V. Assche, J. Cardinal and N. J. Cerf (2004), {\it Reconciliation of a quantum-distributed Gaussian key}, In Proceedings of IEEE Transactions on Information Theory, 50(2), pp. 394-400.
\bibitem{IR1}
F. Grosshans, N. J. Cerf, J. Wenger, R. Tualle-Brouri and P. Grangier (2003), {\it Virtual entanglement and reconciliation protocols for quantum cryptography with continuous variables}, Quantum Inf. Comput. 3, pp. 535-552.
\bibitem{IR4}
X. Q. Jiang, P. Huang, D. Huang, D. Lin and G. Zeng (2017), {\it Secret information reconciliation based on punctured low-density parity-check codes for continuous-variable quantum key distribution}, Phys. Rev. A, 95(2), 022318.
\bibitem{PA1}
C. H. Bennett, G. Brassard, C. Cr\'{e}peau and U. M. Maurer (1995), {\it Generalized privacy amplification}, In Proceedings of IEEE Transactions on Information Theory, 41(6), pp. 1915-1923.
\bibitem{PA2}
D. Deutsch, A. Ekert, R. Jozsa, C. Macchiavello, S. Popescu and A. Sanpera (1996), {\it Quantum privacy amplification and the security of quantum cryptography over noisy channels}, Phys. Rev. Lett. 77(13), 2818.
\bibitem{slice1}
M. Bloch, A. Thangaraj and S. W. McLaughlin (2005), {\it Efficient reconciliation of correlated continuous random variables using LDPC codes}, arXiv preprint cs/0509041.
\bibitem{slice2}
J. Lodewyck, M. Bloch, R. Garc\'{\i}a-Patr\'{o}n, S. Fossier, E. Karpov, E. Diamanti, T. Debuisschert, N. J. Cerf, R. Tualle-Brouri, S. W. McLaughlin and P. Grangier (2007). {\it Quantum key distribution over 25 km with an all-fiber continuous-variable system}, Phys. Rev. A, 76(4), 042305.
\bibitem{MD1}
A. Leverrier, R. All\'{e}aume, J. Boutros, G. Z\'{e}mor and P. Grangier (2008), {\it Multidimensional reconciliation for a continuous-variable quantum key distribution}, Phys. Rev. A, 77(4), 042325.
\bibitem{MD2}
P. Jouguet, S. Kunz-Jacques, A. Leverrier, P. Grangier and E. Diamanti (2013), {\it Experimental demonstration of long-distance continuous-variable quantum key distribution}, Nat. Photon. 7(5), pp. 378-381.
\bibitem{METLDPC}
T. Richardson and R. Urbanke (2002), {\it Multi-edge type LDPC codes}, In Workshop honoring Prof. Bob McEliece on his 60th birthday, California Institute of Technology, Pasadena, California, pp. 24-25.
\bibitem{CODE}
T. Richardson and R. Urbanke (2008), {\it Modern coding theory}, Cambridge University Press (New York).
\bibitem{LDPC}
R. Gallager (1962), {\it Low-density parity-check codes}, In Proceedings of IRE Transactions on Information Theory, 8(1), pp. 21-28.
\bibitem{DVSP}
D. Elkouss, J. Martinez-Mateo and V. Martin (2011), {\it Information reconciliation for quantum key distribution}, Quantum Inf. Comput. 11(3), pp. 226-238.
\bibitem{SP}
D. Elkouss, J. Martinez-Mateo and V. Martin (2010), {\it Secure rate-adaptive reconciliation}, In Proceedings of Information Theory and its Applications, pp. 179-184.
\bibitem{SecSP}
D. Elkouss, J. Martinez-Mateo and V. Martin (2013), {\it Analysis of a rate-adaptive reconciliation protocol and the effect of leakage on the secret key rate}, Phys. Rev. A, 87(4), 042334.
\bibitem{Blind}
J. Martinez-Mateo, D. Elkouss and V. Martin (2012), {\it Blind Reconciliation}, Quantum Inf. Comput. 12(9\&10), pp. 0791-0812.
\bibitem{HIGHspeed}
A. R. Dixon and H. Sato (2014), {\it High speed and adaptable error correction for megabit/s rate quantum key distribution}, Scientific reports 4, 7275.
\bibitem{SymBlind}
E. O. Kiktenko, A. S. Trushechkin, C. C. W. Lim, Y. V. Kurochkin and A. K. Fedorov (2016), {\it Symmetric blind information reconciliation for quantum key distribution}, arXiv preprint arXiv:1612.03673.
\bibitem{RR}
F. Grosshans, G. Van Assche, J. Wenger, R. Brouri, N. J. Cerf and P. Grangier (2003), {\it Quantum key distribution using gaussian-modulated coherent states}, Nature, 421(6920), pp. 238-241.
\bibitem{QRNG}
A. Stefanov, N. Gisin, O. Guinnard, L. Guinnard and H. Zbinden (2000), {\it Optical quantum random number generator}, J. Mod. Optic. 47(4), pp. 595-598.
\bibitem{DE}
T. J. Richardson and R. L. Urbanke (2001), {\it The capacity of low-density parity-check codes under message-passing decoding}, In Proceedings of IEEE Transactions on Information Theory, 47(2), pp. 599-618.
\bibitem{BIBD}
B. Ammar, B. Honary, Y. Kou, J. Xu and S. Lin (2004), {\it Construction of low-density parity-check codes based on balanced incomplete block designs}, In Proceedings of IEEE Transactions on Information Theory, 50(6), pp. 1257-1269.
\bibitem{PEG}
X. Y. Hu, E. Eleftheriou and D. M. Arnold (2005), {\it Regular and irregular progressive edge-growth tanner graphs}, In Proceedings of IEEE Transactions on Information Theory, 51(1), pp. 386-398.
\bibitem{HIGHBIT}
P. Jouguet, D. Elkouss and S. Kunz-Jacques (2014), {\it High-bit-rate continuous-variable quantum key distribution}, Phys. Rev. A, 90(4), 042329.
\bibitem{NONBINLDPC}
C. Pacher, J. Martinez-Mateo, J. Duhme, T. Gehring and F. Furrer (2016), {\it Information Reconciliation for Continuous-Variable Quantum Key Distribution using Non-Binary Low-Density Parity-Check Codes}, arXiv preprint arXiv:1602.09140.

\end{thebibliography}
\end{document}